\documentclass[12pt]{iopart}

\usepackage{amsfonts}
\usepackage{amsmath}
\usepackage{bm}
\usepackage{graphicx}
\usepackage{xcolor}
\usepackage{color}
\usepackage{amssymb}
\usepackage{mwe}
\usepackage{caption}
\usepackage{subcaption}
\usepackage{indentfirst}
\usepackage{booktabs}
\usepackage{listings}
\usepackage{tabularx,array}
\usepackage{multirow}
\usepackage{hyperref}

\newcommand{\comm}[0]{\textcolor{black}}

\usepackage{cite}

\begin{document}
	
	\title{\comm{An Exact Turbulence Law For the Fluid Description of Fusion Edge Plasmas}}

	\author{L. Scarivaglione, and S. Servidio}
	\address{Dipartimento di Fisica, Universit\`a della Calabria, I-87036 Cosenza, Italy}

	\ead{luisa.scarivaglione@unical.it}
	\vspace{10pt}
	\begin{indented}
		\item[] \today
	\end{indented}

\begin{abstract}
\comm{
	Understanding turbulence via simplified fluid models is crucial for optimizing magnetic confinement in tokamak devices.
}\comm{
	In this work, we propose a novel high-order turbulence law that describes the turbulent cascade at the edges of fusion plasmas, namely valid within the Scrape-off Layer (SOL), in the framework of the Braginskii fluid model.
}
Using the Yaglom-Monin approach, we derive an exact relation characterizing density fluctuations in these strongly magnetized systems. 
\comm{
	We obtain a third-order von K\'arm\'an-Howarth equation in increment form for the case of electrostatic Braginskii model, applied to a decaying turbulence regime.
}
\comm{
	The new {\it Yaglom-Braginskii} law is validated through direct numerical simulations within a reduced (two-dimensional) model.
}
Our analysis reveals that the plasma dynamics obey the cross-scale balance, exhibiting a well-defined inertial range of turbulence. This third-order law can provide an accurate measure of the cascade rate of density fluctuations in the scrape-off layer of laboratory plasmas.
	\end{abstract}

	%
	\vspace{2pc}
	
	\section{Introduction}
	\label{intro}
	Turbulence in tokamak plasmas plays a critical role in determining the behavior and performance of fusion reactors \cite{Scott2007, Conway08, KrasheninnikovEA2008, Singh2020}. Understanding turbulent plasma dynamics at the edge of fusion devices is therefore essential for achieving optimal conditions in energy production from nuclear fusion power plants.
	Plasma instabilities and nonlinearities can drive the radial transport of matter and heat from the core to the outer region \cite{Hidalgo1995, Endler1999, LoarteEA07, BoedoEA09, AngioniEA09, Krasheninnikov2020_book}, known as the Scrape-Off Layer (SOL), giving rise to plasma filaments or ``blobs.'' These structures, frequently observed in experiments \cite{DevynckEA06, NoldEA10, KallmanEA10, SechrestEA11, BoedoEA14, Grenfell2020, Garcia2021}, are characterized by a high-density (and high-temperature) core and exhibit a cross-field two-dimensional (2D) mushroom-like shape with an elongated tail.
\comm{
	The interaction between the plasma and the machine wall can be detrimental \cite{HassaneinEA94, RubelEA13}. Plasma blobs can erode the plasma-facing components \cite{ImranEA24}, leading to dilution of the plasma core because of heavy impurities \cite{Angioni21, KorzuevaEA25}.
}	
	 The elongated ``bursts'' represent significant losses of density and momentum from the central (hot) region to the colder edges \cite{Garcia_2009}. Their prevalence is evident from statistical 	analyses of density fluctuations, which reveal a marked deviation of the probability distribution functions from classical Gaussian statistics. Instead, the distributions display positive skewness and a shape indicative of intermittent turbulence \cite{AntarEA01, GarciaEA07, SladkomedovaEA2023}.

	In the tokamak turbulence scenario, several studies reveal that the basic mechanism responsible for the radial transport of blobs is due to a gradient of the magnetic field \cite{D'IppolitoEA02, KrasheninnikovEA04}, whose strength is inversely proportional to the major radius of the torus. This causes: (1) a drift in the curvature of the particles which induces a charge imbalance, (2) a subsequent polarization in the poloidal direction, and (3) radial advection towards the walls. These effects lead to a radial loss of matter with a subsequent waist of the confinement properties \cite{Lawson_1957, Garbet_2006}, being the particles advected across the magnetic field lines due to the electric drift velocity $\mathbf{u}_E = \frac{\mathbf{E} \times \mathbf{B}}{B^2}$. 
\comm{
	Over the years, numerous edge and SOL turbulence models have been developed to study anomalous transport and complex plasma dynamics near the X-point, offering critical insights into power exhaust. These range from high-fidelity (gyro-)kinetic models (5D/6D) to reduced fluid transport models (0D) \cite{FalessiEA18, SchwanderEA24, StegmeirEA24}. Gyrokinetic codes such as COGENT \cite{DorfEA22}, GENE-X \cite{MichelsEA22}, GKEYLL \cite{BernardEA22}, PICLS \cite{BottinoEA25}, and XGC1 \cite{KuEA09} accurately capture microinstabilities that govern processes such as the ion and electron heat transport \cite{XuEA22}.
}

\comm{
	Turbulent coherent, long-lived structures are evident from statistical analyses of density fluctuations. These structures, commonly known as ``plasma blobs'', cause a clear deviation from classical Gaussian statistics. The distributions, indeed, exhibit  
}  	
    positive skewness and shapes characteristic of intermittent turbulence \cite{GarciaEA07, DIppolitoEA11, ChurchillEA17, StegmeirEA19, BernardEA22, SladkomedovaEA2023, AntarEA01}. Further insights have been gained using the so-called $k$-$\epsilon$ (a more comprehensive diagnostic for blob generation than skewness alone) which has provided a new understanding of turbulence spreading rates \cite{ManzEA15}. Given the coherence of the above properties of SOL turbulence, it is essential to establish a theoretical framework that encompasses not only energy distributions (linked to second-order statistics) but also higher-order turbulence analysis, such as third-order laws describing energy flux across scales. Our work extends beyond conventional statistical analyses by applying Kolmogorov’s exact ‘4/5’ law in increment space \cite{KOLMOGOROV_41a, KOLMOGOROV_41b, Frisch1995}, - a cornerstone of turbulence theory that provides a universal, measure of energy transfer.

	This exact relation connects the third-order longitudinal velocity structure function to the energy dissipation rate within the inertial range. Analogously, a so-called ``Yaglom-Monin'' law \cite{Monin1971} can be derived for the third-order mixed structure functions of a passive tracer advected by a turbulent velocity field. This law provides an estimate of the turbulent mixing of the passive tracer in a steady-state, homogeneous flow. The general expression for the ``mixed'' third-order unction is given by:
	\begin{equation}
		\label{Monin_Yaglom-law}
		\langle \delta u_r (\delta \varphi)^2 \rangle = - \frac{4}{3} \varepsilon r, 
	\end{equation}
	where $\varphi({\bf x}, t)$ represents the passive scalar field, ${\bf u}({\bf x}, t)$ is the velocity field, and $\delta$ denotes the increments, defined as $\delta f = f({\bf x}+{\bf r}, t) - f({\bf x}, t)$ (with $f$ being any stochastic variable). Here, $\delta u_r = \delta {\bf u} \cdot {\hat {\bf r}}$ is the longitudinal velocity increment, i.e., the component of the velocity along the increment direction. The angle brackets $\langle \cdot \rangle$ denote an ensemble average over multiple realizations, which, in accordance with the ergodic theorem, is often replaced by a volume average when the characteristic large-scale lengths are smaller than the system size \cite{Lesieur1987}.  Analyzing turbulent fluctuations through the statistical properties of spatial increments offers an elegant and precise methodology: Eq. (\ref{Monin_Yaglom-law}) is valid exclusively within the inertial range, under the assumptions of steady-state, homogeneous, and isotropic conditions. In this regime, $\varepsilon$ represents the turbulent dissipation rate of the scalar field.
\comm{
	The structure function analysis is widely used in turbulence. For a given stochastic $f$, the second order structure function is defined as $S_2(r)=\langle |f(x+r)-f(x)|^2\rangle (\equiv \langle \delta f^2 \rangle)$. This statistical object characterizes field fluctuations and provides insights on how these fluctuations are distributed across different scales, manifesting typical self-similarity in the inertial range, where a power law scaling is observed, i. e $S_2(r) \sim r^{\alpha}$. A formal definition of this quantity will be found in Section 4.
}

\comm{
	Relations analogous to \ref{Monin_Yaglom-law} 
} 
    have been derived in magnetohydrodynamics (MHD) and serve as a criterion for evaluating the presence and extent of the inertial range in turbulent cascades \cite{SorrisoEA07, ServidioEA08, LepretiEA09, WuEA2022, Wang2022_book, PecoraEA2023, IvanovEA2024}. 
	\comm{
		Variations of this law have also been established for compressible cases \cite{GaltierEA2011, AndresEA2018}, in multi-fluid and hybrid electromagnetic models \cite{HellingerEA18}, 
		}
		and local coarse-grained analogs have been explored and applied to fluid turbulence \cite{ServidioEA2022}. For a comprehensive review of the classical turbulence law and its applications, see Ref. \cite{MarinoEA2023}.

	\comm{
	Inspired by the above well-established exact theorems of hydrodynamic turbulence, the scope of the present work focuses on advancing higher-order turbulence analysis in electrostatic regimes – an approach that remains relatively poorly explored in fusion research. The novelty lies not in the simulation methodology itself (which builds upon well-established and relatively simple techniques), but rather in applying refined theoretical tools to directly evaluate the turbulent cascade rate of density fluctuations in the edge-SOL region. The reduced model presented here primarily serves as a benchmark for validating this new diagnostic approach, which could be applied to more sophisticated simulations and experimental data in future work.
}

\comm{
	Here we derive an exact third-order turbulence law, based on the Yaglom-Monin procedure of spatial increments, for the Braginskii approximation of two-fluid turbulence. We adopt a SOL model that is electrostatic, two-dimensional, that disregards parallel dynamics and assumes a cold ion limit.
}

\comm{
	We establish a relation that characterizes the evolution and distribution of density fluctuations in the compressible electrostatic regime in the form of a Yaglom-like relation for the third-order mixed moment--involving the particle number density as a scalar and the $\mathbf{E} \times \mathbf{B}$ drift velocity-- that could provide significant insights into turbulent transport at the edges of tokamaks.
}
We propose a new law by deriving a scale-by-scale balance equation for density fluctuations, incorporating the mixed third-order moment. 
\comm{ 
	Our theoretical result has been validated using direct numerical simulations of Braginskii turbulence in a reduced 2D system.
}
	
The paper is organized as follows. In Section \ref{sec:model}, we briefly outline the governing equations of our electrostatic numerical model. Section \ref{sec:New third-order equation} details the derivation of the third-order law for electrostatic turbulence. Section \ref{sec:results} presents the key findings of our study. Finally, discussions and conclusions are provided in the last section.

\section{The Turbulence Model}
\label{sec:model}
To describe the plasma dynamics at the edge of fusion devices, we employ a model based on the drift-reduced Braginskii equations \cite{Braginskii_1965}, which effectively capture the evolution of macroscopic quantities in the edge boundary layers \cite{HasegawaWakatani1983, RicciEA2012, WiesenEA2015, TamainEA2016, MadsenEA2016, ThrysoeEA2018, RivaEA2019, StegmeirEA2019, PoulsenEA2020, GiacominEA2022, SchwanderEA24}. In this region, the plasma is relatively cold, densities remain high, and the collisionality justifies the use of a reduced fluid description \cite{SchwanderEA24, Chen_1984, SimakovEA04}. The primary quantities are derived by integrating the distribution function in velocity space.

The model relies on the quasi-neutrality assumption, $n_e \simeq n_i \simeq n$, where $n$ represents the plasma density, and the subscripts ``e'' and ``i'' denote electrons and ions, respectively. Under electrostatic conditions, the electric field $\mathbf{E}$ is given by $\mathbf{E} = -\mathbf{\nabla} \phi$, where $\phi$ is the electrostatic potential. For simplicity, and in line with the arguments presented in Ref.s~\cite{GarciaEA05, FundamenskiEA07, MilitelloEA12}, we adopt the cold ions approximation, $T_i = 0$, and assume isotropic pressure, reducing the pressure tensor to $\mathbf{P} = p \mathbf{I}$, where $p = n T$ and $T$ is the electron temperature. Under these assumptions, the drift-reduced Braginskii equations take the form:
	\begin{gather}
		\label{continuity_eq}
		\frac{dn}{dt} + n C(\phi) - C(nT) = \nu_n \nabla^2 n,  \\
		\label{energy_eq}
		\frac{dT}{dt} - \frac{7T}{3} C(T) - \frac{2T^2}{3n} C(n) + \frac{2T}{3} C(\phi) = \nu_T \nabla^2 T, \\
		\label{vorticity_time}
		\frac{d \Omega}{dt} - C(nT) = \nu_\Omega \nabla^2 \Omega, \\
		\label{vorticity_eq}
		\Omega = \nabla^2 \phi,
	\end{gather}
\comm{
	the vorticity $\Omega$ is derived from the difference between the ion and electron continuity equations, using the Boussinesq approximation.
} 
    The total time derivative is defined as $\frac{d}{dt} = \frac{\partial}{\partial t} + \mathbf{u} \cdot \mathbf{\nabla}$, combining the partial time derivative with the advection term due to the $\mathbf{E} \times \mathbf{B}$ drift velocity. The operator $C(\bullet)$ accounts for the compressibility of the system, arising from magnetic curvature effects \cite{Garcia_2003}, as detailed below.

	In the above set of equations, the velocity corresponds to the $\mathbf{E} \times \mathbf{B}$ drift, given by $\mathbf{u}_E = \frac{\mathbf{E} \times \mathbf{B}}{B^2}$. The model employs local slab coordinates, where the Cartesian axes $x$, $y$, and $z$ represent the radial, poloidal, and toroidal coordinates of the device, respectively. The terms of the form ``$\nu_f \nabla^2$'' represent diffusive processes, with $\nu_f$ denoting dissipation coefficients \cite{GarciaEA05}.
\comm{
		The magnetic field is oriented along the $\hat{\bf{z}}$ direction and is expressed as $\mathbf{B} = B(x) \hat{\bf{z}} = \frac{\mathcal{B}}{1 + \epsilon + \zeta x} \hat{\bf{z}},$ with $\mathcal{B}$ a characteristic magnetic field strength, and  $\epsilon$ is the inverse aspect ratio. The radial gradient is instead  governed by the parameter $\zeta$.
	}
	The curvature operator appearing in Eqs.~(\ref{continuity_eq})-(\ref{vorticity_eq}) is defined as $C = -\zeta \frac{\partial}{\partial y}$. We assume $\mathbf{u}_E$ to be the dominant contribution to the plasma motion, allowing the vorticity $\Omega$ to be expressed as the curl of the electric drift in the plane perpendicular to $\mathbf{B}$, i.e., $\Omega = \mathbf{\hat{z}} \cdot (\nabla \times {\mathbf{u_E}}) = \frac{\nabla^2{\phi}}{B}$ \cite{MilitelloEA12}. 
	
	The equations~(\ref{continuity_eq})-(\ref{vorticity_eq}) are normalized using the Bohm normalization, defined as:
	\begin{equation}
		\omega_{ci}t \rightarrow t, \quad \frac{\mathbf{x}}{\rho_s} \rightarrow \mathbf{x}, \quad  \frac{q \phi}{\mathcal{T}} \rightarrow \phi, \quad \frac{n}{\mathcal{N}} \rightarrow n,  \quad  \frac{T}{\mathcal{T}} \rightarrow T,
		\label{eq:bhom}
	\end{equation}
	where $\omega_{ci} = q\mathcal{B}/m_i$ is the ion gyrofrequency, $\rho_s = c_s/\omega_{ci}$ is the hybrid thermal gyroradius, $c_s = (\mathcal{T}/m_i)^{1/2}$ is the ion sound speed, and $\mathcal{N}$ and $\mathcal{T}$ are characteristic values of the particle density and electron temperature, respectively. For further details on the model, see Ref.~\cite{ScarivaglioneEA23}.

\comm{
	The simplified numerical model adopted for this analysis recalls other existing (more refined) codes in the context of edge-SOL fluid simulations, such as HESEL \cite{RasmussenEA15}, GBS \cite{RicciEA12}, GRILLIX \cite{StegmeirEA18} or STORM \cite{AhmedEA25}. The background assumptions in Eq.s~(\ref{continuity_eq})-(\ref{vorticity_eq}) allow us to significantly reduce the computational costs and the complexity of the code. Similarly, the cold ion assumption provides further simplification and is valid under the assumption of high-collisionality, as demonstrated by \cite{FundamenskiEA07}.  We intentionally start from a simpler approach to keep things straightforward to validate the analysis.
}

\comm{
	The diffusive (viscous) terms in Eq.s~(\ref{continuity_eq})-(\ref{vorticity_eq}) serve primarily to maintain solution regularity and numerical stability, rather than modeling any specific physical phenomenon. In laboratory plasmas, diffusive processes emerge from various mechanisms such as ionization and collisional transport - the latter, including frictional effects, has been extensively discussed in previous works \cite{MadsenEA2016}. Plasma diffusive effects can establish strong nonlinear density-temperature relationships that our simplified model does not capture. However, we note that, in the case of high Reynolds number turbulence, plasmas might exhibit finite turbulence transfer (dissipation) rates that become independent of the specific viscous coefficients \cite{MininniEA09}.
}

\section{The Yaglom-Braginskii Third-Order Law}
\label{sec:New third-order equation}
The properties of electrostatic turbulence can be examined by studying the scale-dependent increments of macroscopic plasma fluctuations. This analysis can be conducted directly in physical space using two-point correlation functions and structure functions, applying the Yaglom-Monin procedure \cite{Monin1971} to the reduced drift-Braginskii equations. This approach has been then analogously applied by Politano and Pouquet to incompressible MHD \cite{Politano_Pouquet1995}.

	Let us consider the plane perpendicular to the local magnetic ﬁeld, and two points in this plane separated by a distance $\mathbf{r}$, such that $\mathbf{x'} = \mathbf{x} + \mathbf{r}$. We now evaluate the Braginskii continuity equation for particle density, Eq~(\ref{continuity_eq}), at both points at a fixed time, obtaining
	\begin{gather}
		\label{n'}
		\frac{ \partial n'}{\partial t} + u'_j \frac{\partial n'}{\partial x'_j}  + n' C'(\phi') - C '(n ' T ') = \nu \frac{\partial ^2}{\partial x'^2_k} n' , \\
		\label{n}
		\frac{ \partial n}{\partial t} + u_j \frac{\partial n}{\partial x_j}  + n C(\phi) - C(nT) = \nu \frac{\partial ^2}{\partial x^2_k} n ,
	\end{gather}
	where the suffix `` $'$ '' refers to quantities evaluated in $\mathbf{x'}$, namely $f'=f(\mathbf{x'})$. 
	As usual in the classical picture of homogeneous turbulence, we can assume that every field, evaluated in one point is independent of any other point in the phase space (independence hypothesis), namely $\frac{\partial n'}{\partial x_i} = \frac{\partial n}{\partial x'_i} = 0 $. 
\comm{
		We can therefore introduce the increments of the type $\delta n = n'-n = n ( \mathbf{x}+\mathbf{r} ) - n(\mathbf{x} )$, and consider the difference between Eq.s~(\ref{n'}) and (\ref{n}) as  
}	
	\begin{gather}
		\frac{ \partial (\delta n)}{\partial t} + u'_j \frac{\partial (\delta n)}{\partial x'_j} + u_j \frac{\partial (\delta n)}{\partial x_j} + n' C'( \phi ') - n C( \phi) - C'( n' T' ) + C( n T ) \notag \\
		=  \nu \left( \frac{\partial ^2}{\partial x'^2_k} + \frac{\partial ^2}{\partial x^2_k} \right) (\delta n). 
	\end{gather}
	After simple manipulation, multiplying by the term $2 \delta n$, and averaging over the volume, we obtain 
	\begin{gather}
		\left \langle \frac{ \partial (\delta n)^2}{\partial t} \right \rangle  = - \left \langle \delta u_j \frac{ \partial (\delta n)^2}{\partial x'_j}\right \rangle  - \left \langle u_j \left( \frac{\partial }{\partial x'_j} + \frac{\partial }{\partial x_j} \right) (\delta n)^2 \right \rangle \notag \\
		- \langle  2 n' \delta n C'( \phi ' ) \rangle +\langle  2 n \delta n C( \phi )  \rangle + \langle 2 \delta n C'( n' T ' ) \rangle \notag \\
		\label{increments}
		- \langle 2 \delta n C( n T  )  \rangle + \left \langle  2  \nu \delta n \left( \frac{\partial ^2}{\partial x'^2_k} + \frac{\partial ^2}{\partial x^2_k} \right) \delta n  \right \rangle.
	\end{gather}
	We assume that the system is locally homogeneous, so that $ \left \langle \frac{\partial }{\partial x'_i} ... \right \rangle = - \left \langle \frac{\partial }{\partial x_i}... \right \rangle = \frac{\partial }{\partial r_i} \langle ... \rangle$ \cite{Politano_Pouquet1995}. 

\comm{
		This assumption, together with introducing the divergence as a source field $\chi=\partial_j u_j$, leads to a straightforward simplification of the latter equation, which becomes
} 
	\begin{gather}
		\frac{ \partial }{\partial t} \langle (\delta n)^2  \rangle = -  \frac{\partial }{\partial r_j}\langle \delta u_j (\delta n)^2  \rangle + \langle \chi ' (\delta n)^2  \rangle  + \langle \chi  (\delta n)^2  \rangle \notag \\
		- \langle  2 n' \delta n C '( \phi ' )  \rangle  + \langle  2 n \delta n  C( \phi )  \rangle + \langle  2 \delta n C'( n' T  ' )  \rangle \notag \\
		\label{eq_avg_1}
		- \langle  2 \delta n C( n T  )  \rangle + 2 \frac{\partial^2}{\partial r_k^2} \langle  \nu (\delta n)^2 \rangle - 4  \langle \nu (\partial_k n)^2 \rangle. 
	\end{gather}
	If one defines $ \bar{\chi} = \frac{1}{2}( \chi ' + \chi )$, as suggested in \cite{GaltierEA2011}, and given the curvature operator $C = -\zeta \frac{\partial}{\partial y}$, the above equation reads 
	\begin{gather}
		\frac{\partial}{\partial t}\left\langle(\delta n)^2\right \rangle\!=\!-\frac{\partial}{\partial r_j}\langle\delta u_j(\delta n)^2\rangle + 2 \langle \bar{\chi} (\delta n)^2  \rangle - \left \langle 2 \zeta n' \delta n \left(-\frac{\partial \phi '}{\partial y '} \right )\right \rangle + \left \langle  2 \zeta n \delta n \left (- \frac{\partial \phi}{\partial y} \right ) \right \rangle \notag \\
		\label{merging}
		-\left \langle 2 \zeta \delta n  \frac{\partial (n' T ')}{\partial y'} \right \rangle + \left \langle 2 \zeta \delta n  \frac{\partial (n T) }{\partial y} \right \rangle +2 \frac{\partial^2}{\partial r_k^2} \left \langle  \nu (\delta n)^2  \right \rangle - 4 \langle \nu (\partial_k n)^2 \rangle.  
	\end{gather}
	In order to have a more compact form, we use the electric field $\mathbf{E} = - \nabla \phi$ and, introducing the pressure gradient field $\mathbf{G} = {\mathbf \nabla} P = {\mathbf \nabla}( n T )$, the final expression for the Yaglom-Braginskii third-order law is 
	\begin{gather}
		\varepsilon = 
		-\frac{1}{4}\frac{\partial}{\partial t}\left\langle(\delta n)^2\right\rangle
		+\frac{1}{2}\left\langle\bar{\chi} (\delta n)^2\right\rangle
		-\frac{1}{2}\zeta\left\langle\delta n\delta( n E_y )\right\rangle
		-\frac{1}{2}\zeta\left\langle\delta n\delta G_y\right\rangle \notag \\
		-\frac{1}{4}\frac{\partial}{\partial r_j}\left\langle\delta u_j (\delta n)^2\right\rangle
		+\frac{1}{2}\frac{\partial^2}{\partial r_k^2}\langle \nu (\delta n)^2\rangle
		\label{final-law}, 
	\end{gather}
	where the cascade rate of the density is
\comm{  
	$\varepsilon = \langle \nu (\partial_k n)^2 \rangle$,
}
	in analogy with the cascade rate of the Yaglom-Monin model of turbulence \cite{Monin1971}.

	The new result expressed by Eq.~(\ref{final-law}) represents a von K\'arm\'an-Howarth equation in increment form for Braginskii turbulence, providing a scale-by-scale budget for the cascade of density fluctuations. In this balance equation, each term corresponds to a distinct physical effect. The first term on the right-hand side ($\frac{\partial}{\partial t}\left\langle(\delta n)^2\right\rangle$) represents the time derivative of the second-order structure function of the density fluctuations, capturing the large-scale temporal variation of the second-order moment. This term quantifies the unsteadiness of the cascade at large scales (and low frequencies), or equivalently, the rate of change of density second-order statistics over time. The term $\left\langle\bar{\chi} (\delta n)^2\right\rangle$ quantifies large-scale inhomogeneity associated with the compressibility of the velocity field. As in compressible versions of the third-order law \cite{GaltierEA2011}, this term can act as a source of the cascade and may become negligible in nearly incompressible regimes, such as in electrostatic edge turbulence. The third term on the right-hand side is linked to the sheared electric field, an effect arising from magnetic field curvature, which can introduce anisotropy in the fluctuations. Similarly, the fourth term, $\zeta\left\langle\delta n\delta G_y\right\rangle$, represents a shear in the pressure balance. 
	
	The most significant contribution is the scale-by-scale Yaglom term, $\frac{\partial}{\partial r_j}\left\langle\delta u_j (\delta n)^2\right\rangle$, which describes the cascade rate of density fluctuations advected by the velocity field. This term represents the nonlinear transfer of density fluctuations across scales, uniquely defining the inertial range of turbulence \cite{Politano1998, ServidioEA08, LepretiEA09, Monin1971, WangEA2022}. This mixed third-order moment is expected to match $\varepsilon$ if a clear separation of scales exists.  Finally, the last term, $\frac{1}{2}\frac{\partial^2}{\partial r_k^2}\langle \nu (\delta n)^2\rangle$, becomes dominant in the viscous regime, where dissipation plays a significant role. This term is expected to prevail at very small scales, where the cascade ultimately terminates.

	It is important to note that, due to homogeneity, taking either an ensemble average or a spatial average (denoted here as $\langle ... \rangle$) implies that all increments depend solely on the lag $r$ \cite{Mccomb1995}. 
\comm{
     Furthermore, since the mean dissipation rate of the density, $\varepsilon = \langle \nu (\partial_k n)^2 \rangle$, involves real-space quantities and a constant dissipation coefficient, it is independent of the lag and, in the Kolmogorov picture, remains constant across all length scales. 
}
The new Yaglom-Braginskii third-order law in Eq.~(\ref{final-law}) provides a comprehensive description of the entire cascade process of density fluctuations, encompassing their generation at large scales, their transfer across scales, and their eventual dissipation. We performed direct numerical simulations of the drift-Braginskii equations described in Eq.(\ref{continuity_eq})-(\ref{vorticity_eq}), in order to verify the validity of the new law, as described in the following section.

	\section{Numerical Simulations of Electrostatic Drift-Braginskii Turbulence}
	\label{sec:results}
	We now present direct numerical simulations of electrostatic drift-Braginskii turbulence, governed by Eqs.~(\ref{continuity_eq})-(\ref{vorticity_eq}). Specifically, these simulations are conducted in a reduced two-dimensional (2D) geometry, perpendicular to the out-of-plane magnetic field $B_z$. The dynamics are confined to the ($x$, $y$) plane, where periodicity is assumed in the $y$-direction, and the primary field gradient is aligned with the $x$-axis. This setup mimics the outward radial direction characteristic of edge turbulence.
	
	The numerical algorithm employs second-order centered finite differences for spatial differentiation and a second-order Runge-Kutta scheme for the temporal evolution of the system. To solve the Poisson equation for the electrostatic potential, a combination of finite differences and spectral methods is used, a common approach for systems with a periodic direction. Further details on the numerical methods and geometry are provided in \cite{ScarivaglioneEA23}. The simulation domain is a rectangular box of size $L_x \times L_y$, discretized using a cell-centered grid with $N_x$ and $N_y$ equidistant points in the $x$- and $y$-directions, respectively. For this analysis, the simulation resolution is $N_x \times N_y = 2048 \times 2048$ mesh points, with a domain size of $L_x \times L_y = 400 \times 400$ in normalized Bohm units [see Eq.~(\ref{eq:bhom})].
\comm{
	A constant dissipation coefficient of $\nu = 3 \times 10^{-3}$ is adopted to ensure numerical stability throughout the simulation.
}
\comm{
	The simulation domain encompasses the edge plasma column, the Last Closed Flux Surface (LCFS), the SOL, and the near-wall region. The radial position of the LCFS is set at the sharp initial gradients of the density and temperature profiles. While this location ($x_0=100$) is arbitrary in our reduced decaying model, it was selected to optimally facilitate the growth of instabilities and the subsequent development of fully turbulent dynamics in the outer region.
}

	We aim to mimic the conditions of a typical discharge, where turbulence emerges at the sharp edge gradients of the SOL. Eq.s~(\ref{continuity_eq})-(\ref{vorticity_eq}) are solved in a decaying turbulence regime, with an initial imposition of sharp gradients in both density and temperature. The equilibrium is then perturbed with minimal random fluctuations. Following an initial transient phase, a quasi-steady-state turbulence is established, which persists for an extended period. During this phase, energy is injected from the large-scale shear of density and temperature and subsequently dissipated at very small scales through the viscous terms in the equations. Furthermore, turbulence generated by large-scale gradients propagates outward in the $x$-direction, forming blob-like structures that continuously leave the domain, resembling the bursty behavior of edge turbulence observed in laboratory plasmas \cite{HidalgoEA03, DevynckEA06, GarciaEA07, DudsonEA08, IonitaEA09, KallmanEA10, SechrestEA11, FasoliEA2010, BoedoEA14}. Within this quasi-homogeneous, steady-state transient, we conduct the primary statistical analysis of the Yaglom-Braginskii law. Notably, our numerical simulations intentionally avoid any external ad-hoc driving mechanisms, ensuring the absence of artificial injection disturbances and enabling a natural validation of the law.
	
	\begin{figure}
		\centering
		\includegraphics[width=0.8\linewidth]{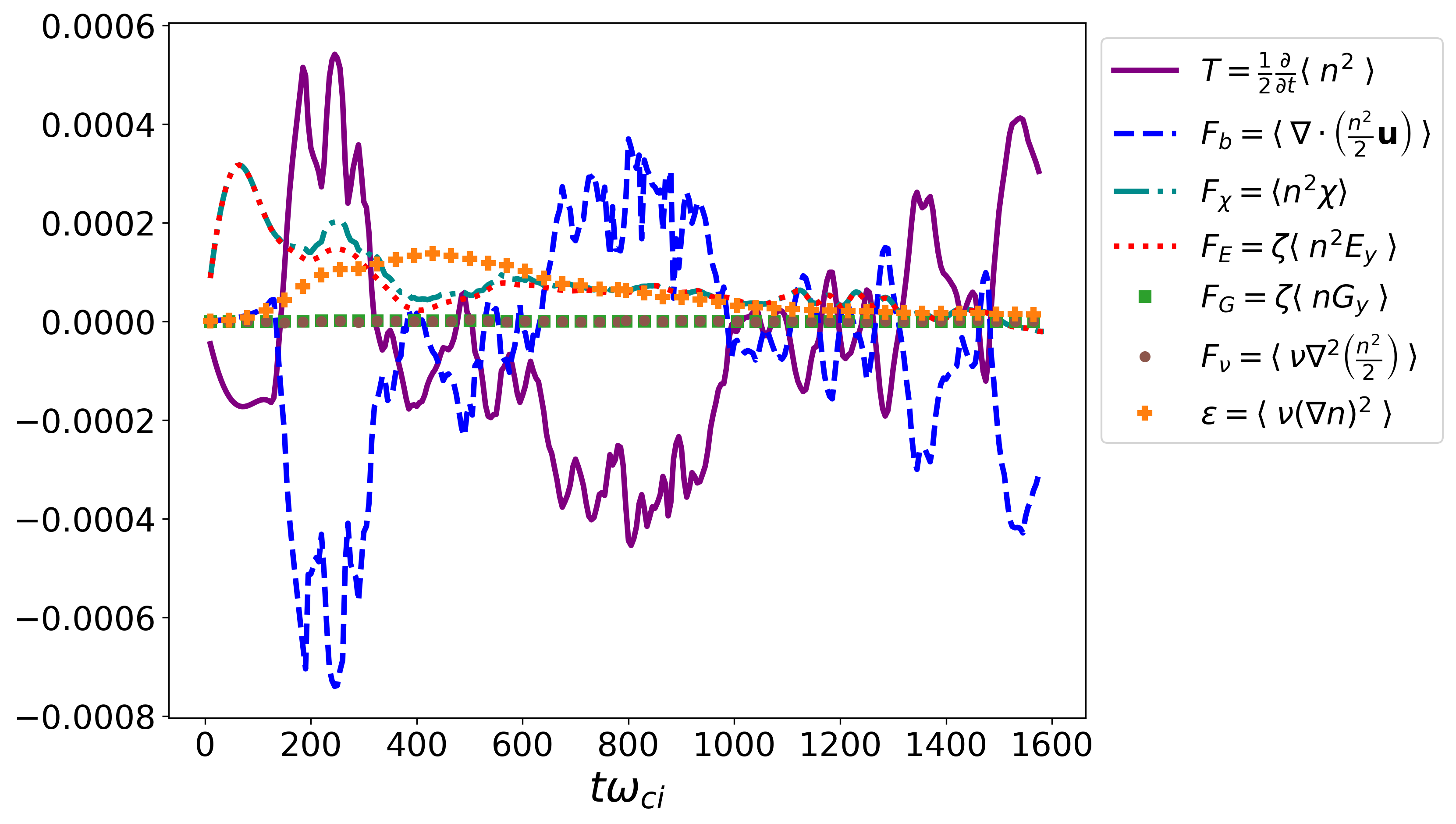}
		\caption{Global budget of the density fluctuations. Different symbols (colors) correspond to the terms in Eq.~(\ref{density_variance}). }
		\label{fig:density_variance}
	\end{figure}

	\subsection{Second-Order Statistics}
	\begin{figure}
		\centering
		\includegraphics[width=0.8\linewidth]{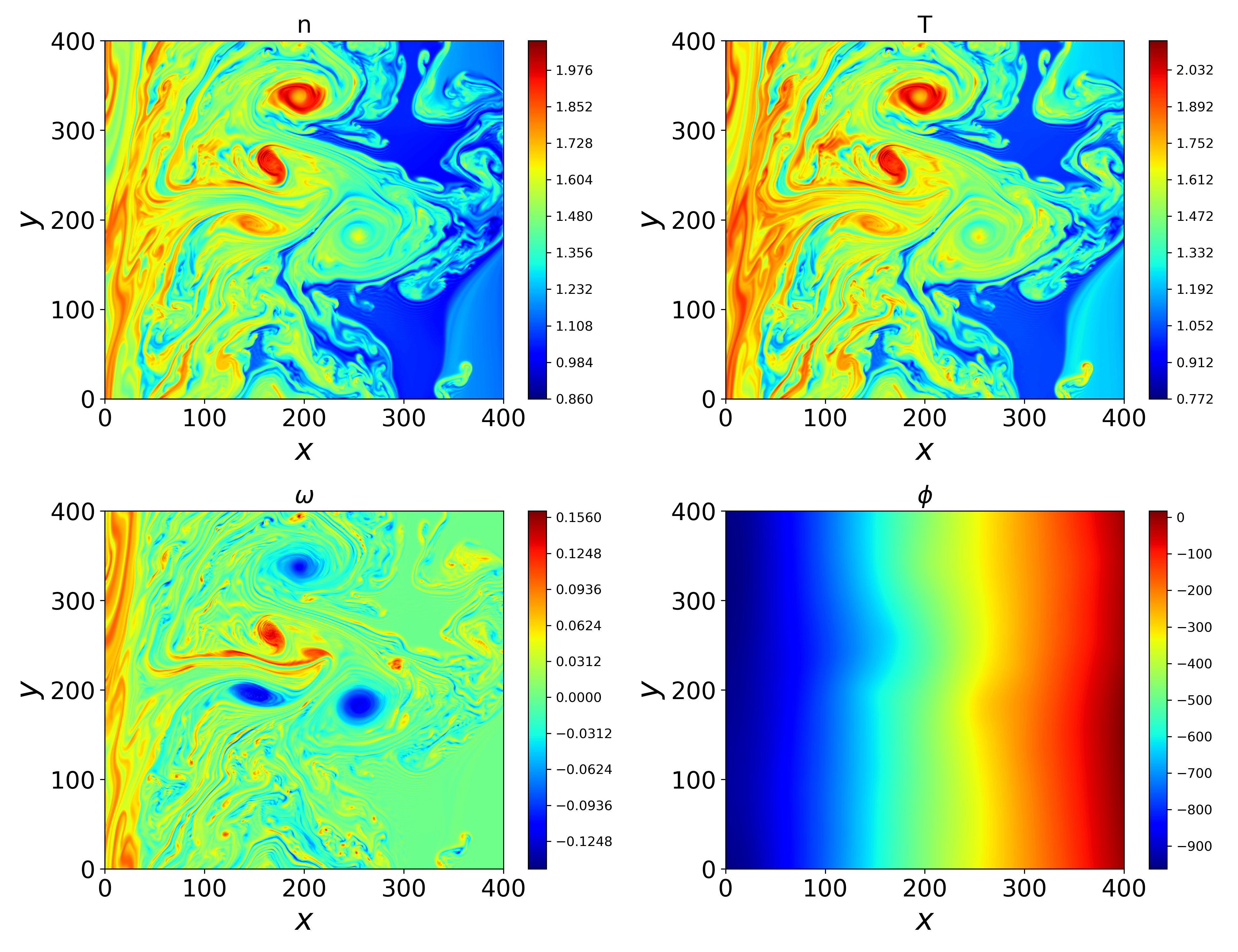}
		\caption{Two-dimensional spatial structure of particle density (top-left), temperature (top-right), vorticity (bottom-left), and potential (bottom-right), at  $t=1400$.}
		\label{fig:fields}
	\end{figure}
	To qualitatively characterize the global behavior of the system, we investigated the temporal evolution of the density variance.  The latter can be obtained by the continuity Eq.~(\ref{continuity_eq}) multiplied by $n$, which after volume averaging becomes
	\begin{gather} 
		\frac{1}{2} \frac{\partial \langle n^2 \rangle}{\partial t}  + \left \langle \nabla \cdot \left( \frac{n^2}{2} \mathbf{u} \right) \right \rangle - \langle n^2 \chi \rangle + \zeta \langle n^2 E_y \rangle + \zeta \left \langle n G_y \right \rangle - \left \langle \nu \nabla^2 \left( \frac{n^2}{2} \right) \right \rangle  +  \langle \nu (\nabla n)^2 \rangle =  \notag \\
		\label{density_variance}
		T + F_b + F_\chi + F_E + F_G + F_\nu + \varepsilon = 0.
	\end{gather}
\comm{
	This global conservation law, typical of turbulence analysis \cite{ManzEA15}, resembles the structure of the Yaglom-Braginskii balance in Eq.~(\ref{final-law}).
}
The values of each term in Eq.~(\ref{density_variance}) are displayed in Figure \ref{fig:density_variance}. The evolution begins with a rapid increase in the electric field term $F_E$ and the divergence term $F_\chi$, driven by shear that injects energy into the system. This is followed by the growth of dissipative terms, signaling the onset of the cascade. The most significant fluctuations arise from the boundary flux term $F_b = \left \langle \nabla \cdot \left( \frac{n^2}{2} \mathbf{u} \right) \right \rangle$, which exhibits strong anti-correlation with the time variation term $T$. These anti-correlated bursts are attributed to events of plasma flux through the open boundaries. For instance, $F_b$ ($T$) peaks around $t = 200$, coinciding with the formation and development of structures across the LCFS, leading to a flux of particle density at the open boundaries, as will be demonstrated later.

	The dissipation (cascade) rate $\varepsilon$ gradually increases after $t=200$, after which the quasi-steady-state is achieved. The peak occurs around $t \sim 450$, coinciding with the formation of the first blob structures in the SOL, which triggers an efficient turbulent cascade toward smaller scales. At this stage, the system is expected to reach a balance between anisotropic energy injection, boundary fluxes at the edges, and the cascade toward small scales. The term $F_\nu$ remains relatively small but non-zero, as it represents the flux of viscous terms through the boundaries. Additionally, we computed the sum of all terms in Eq.~(\ref{density_variance}) and confirmed that the balance is achieved, with a small error of (at most) 7\%, that rapidly converges to zero at longer times. The latter is consistent with the use of finite differences and the fact that during the initial stages of the simulation, the system is still reacting to the initial conditions.

	The numerical code reproduces the turbulent dynamics of the outer region of fusion devices, as illustrated in Fig.~(\ref{fig:fields}), which displays 2D shaded contours of the density $n$, temperature $T$, vorticity $\omega$, and electrostatic potential $\phi$. Prominent high-density and high-temperature structures are evident, originating near the LCFS at $x=100$, and propagating into the SOL for $x>100$, eventually reaching the near-wall region. These structures exhibit a characteristic mushroom-like shape with elongated tails, consistent with observations in fusion devices \cite{FurnoEA08, MullerEA09, DIppolitoEA11, DecristoforoEA20, BisaiEA22}. In contrast, the inner region ($x<50$) is dominated by plumes that later evolve into blob-like structures. The electrostatic potential $\phi$ exhibits a gradient correlated with the presence of polarization and the associated $\mathbf{E} \times \mathbf{B}$ drift.

	\begin{figure}
		\centering
		\includegraphics[width=0.7\linewidth]{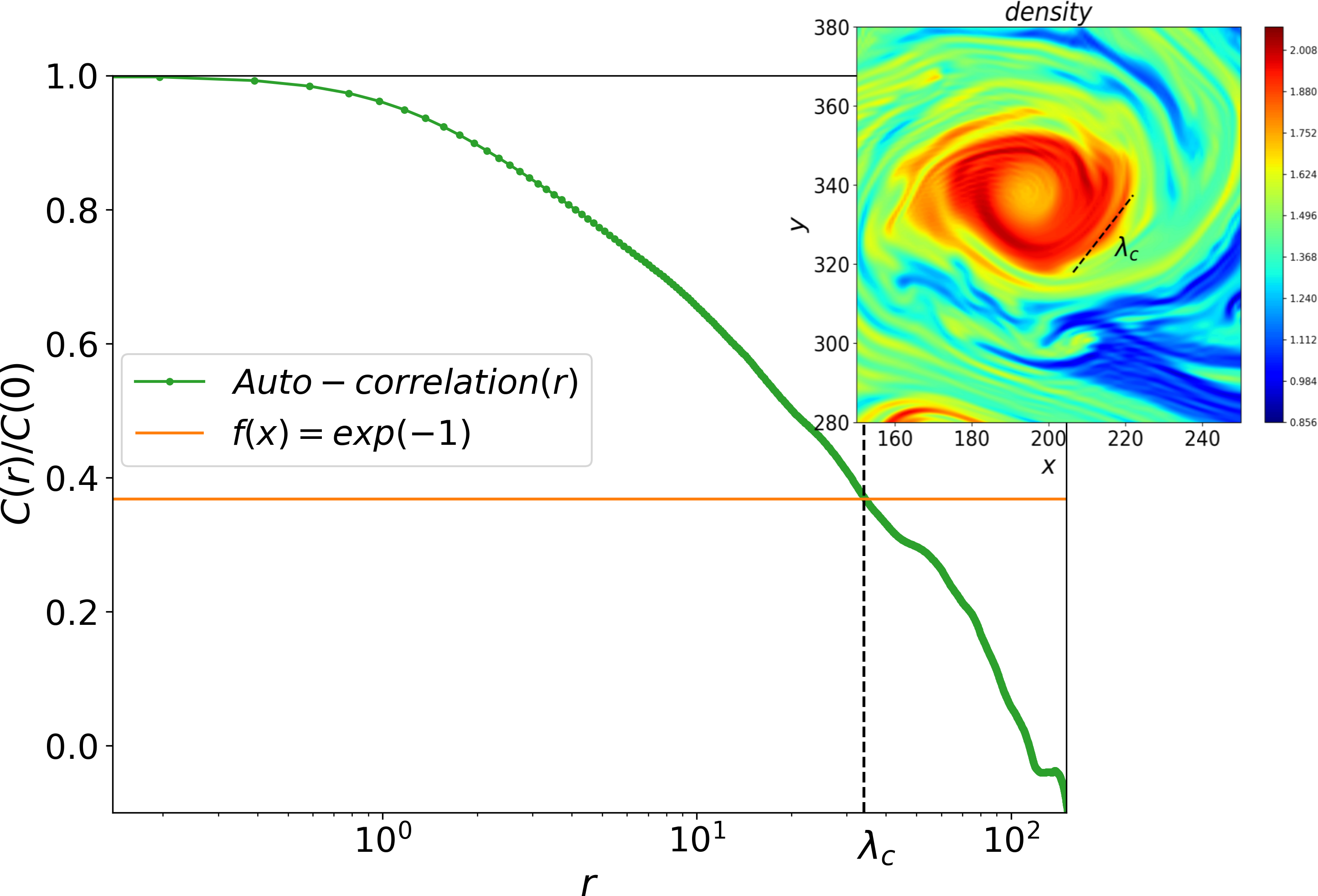}
		\caption{Normalized auto-correlation function $C(r)/C(0)$, evaluated at $t = 1400$, as a function of the increment $r$. The correlation length $\lambda_c$, defined as the $e$-folding distance of $C(r)/C(0)$, is indicated by a vertical dashed line. In the top-right inset, the density contours are displayed, featuring a characteristic blob structure.}
		\label{fig:Correlation}
	\end{figure}

	Before focusing on the high-order exact law, it is essential to measure and understand the characteristic scales of turbulence through classical statistical analysis. To this end, we investigate the properties of turbulent density fluctuations by analyzing increments at various spatial lags. Specifically, we evaluate the auto-correlation function and the second-order structure function of the density field. We examine these statistical quantities as functions of the spatial lag $\mathbf{r}$. This analysis is conducted at a specific time during the steady-state evolution ($t = 1400$) within a restricted domain characterized by fully developed homogeneous turbulence (from $x = 80$ to $x = 300$, in the SOL), thereby excluding the inner and outer boundaries. For consistency, we also performed the analysis at different times and under varying initial conditions, obtaining similar results (see later).

	The auto-correlation function of the density fluctuations is defined as
	\begin{equation}
		C( {\bf r} ) = \langle \tilde n( \mathbf{x} + \mathbf{r} ) \tilde n( \mathbf{x} ) \rangle = \frac{1}{V} \int \tilde n( \mathbf{x} + \mathbf{r} ) \tilde n( \mathbf{x} ) d \mathbf{x}, 
	\end{equation}
	where $\tilde n = n - \langle n \rangle$ represents the fluctuations around the mean density. To extract isotropic information about the characteristic length scales, we average the auto-correlation function along $r_x$ and $r_y$, yielding $C(r) = [C(r_x)+C(r_y)]/2$ \cite{SorrisoEA02} (with $r_x=r_y\equiv r$). This isotropic correlation function, normalized to the variance $C(0)$, is shown in Figure \ref{fig:Correlation} and exhibits the typical profile of stochastic fluctuations in turbulence \cite{Frisch1995}. The normalized function $C(r)/C(0)$ provides insight into the typical size of energy-containing eddies, $\lambda_c$, which can be determined either by integrating $C(r)/C(0)$ or empirically as its $e$-folding length. For these numerical experiments, as illustrated in the same figure, we measure 
\comm{ 
		$\lambda_c \simeq 34$. 
}	
	This scale, which qualitatively corresponds to the size of bursty blobs \cite{ScarivaglioneEA23, ServidioEA08}, is significantly smaller than the domain size, $L_x = L_y = 400$, indicating a state of relatively homogeneous turbulence. In the top-right panel of Figure \ref{fig:Correlation}, we display the 2D turbulent density pattern, highlighting the scale of $\lambda_c$.
	
	\begin{figure}
		\centering
		\includegraphics[width=1.\linewidth]{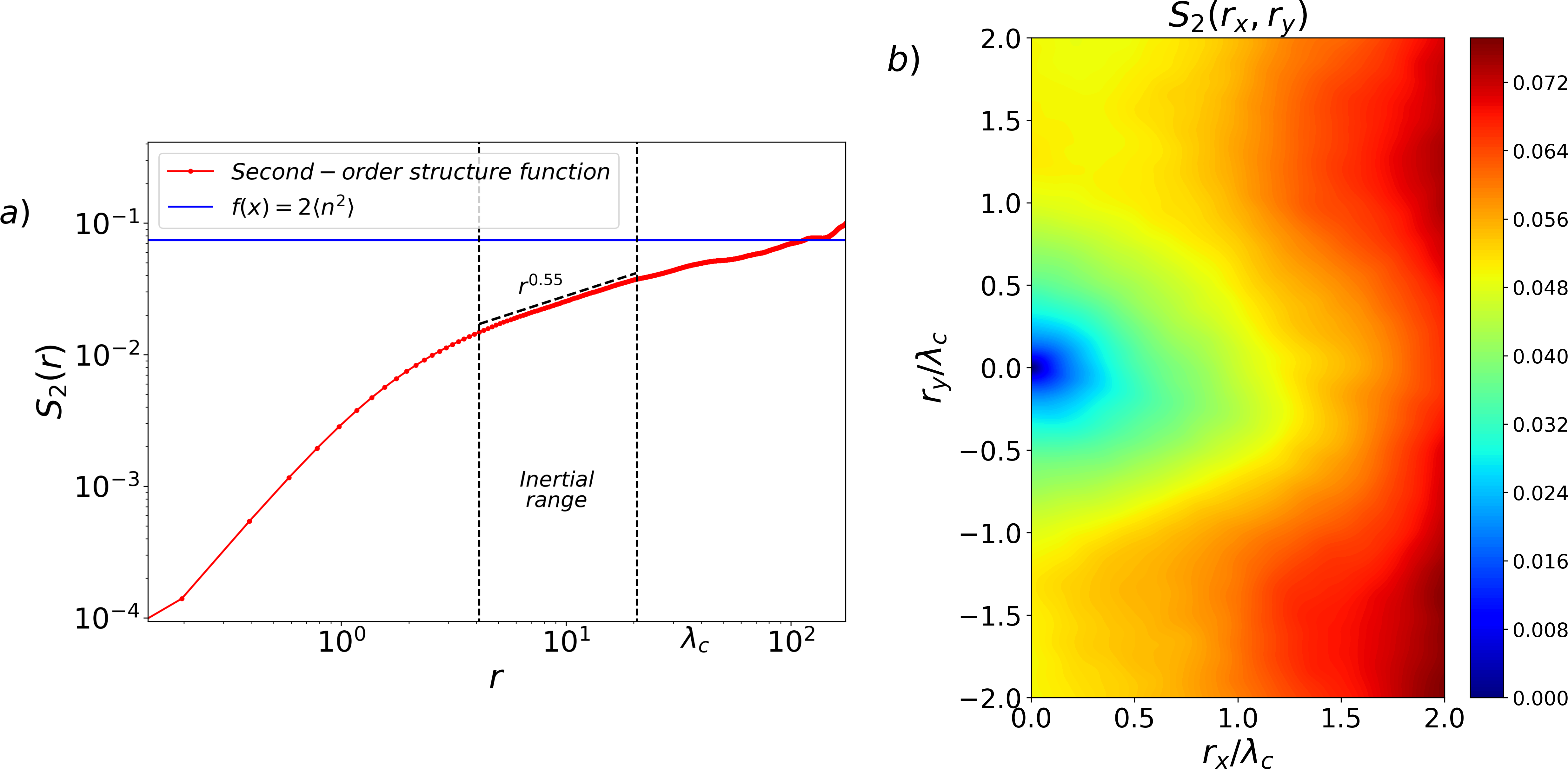}
		\caption{(a) Isotropic second-order structure function of $n$ as a function of the  increment $r$. The correlation length $\lambda_c$ is indicated on the $x$-axis. For scales larger than $\lambda_c$, $S_2(r)$ converges to a plateau, as expected, reaching a value equal to twice the density variance (blue horizontal line). At very small scales ($r \rightarrow 0$), dissipation effects dominate. Within the inertial range, $S_2(r)$ exhibits a power-law scaling with a slope of approximately $\sim 0.55$. (b) Two-dimensional representation of the $S_2({\mathbf r}$), as a function of the increments normalized to the correlation lengths.}
		\label{fig:S2}
	\end{figure}

	In conjunction with the auto-correlation function, we evaluated the second-order structure function of the density, defined as
	\begin{gather}
		S_2( {\bf r} ) = \langle (n( \mathbf{x} + \mathbf{r} )  - n ( \mathbf{x} ))^2 \rangle = \langle (\delta n(\mathbf{r} ))^2 \rangle
		= 2\langle {\tilde n}^2 \rangle - 2 C({\bf r}).
		\label{eq:s2}
	\end{gather}
	The isotropic structure function is displayed in Figure \ref{fig:S2}-(a) as a function of the spatial lag $r$. For scales larger than $\lambda_c$, $S_2$ converges to a plateau, as expected, indicating the absence of correlations. At this scale, $S_2$ reaches twice the value of the variance $\langle {\tilde n}^2\rangle$, consistent with Eq.~(\ref{eq:s2}). At small scales ($r \rightarrow 0$), dissipative effects dominate the dynamics. In the intermediate range, we observe an inertial range of turbulence, where the structure function follows a power-law behavior, $S_2(r) \sim r^{\alpha}$. In this regime, we measured $\alpha \sim 0.55$, which deviates slightly from the Kolmogorov K41 scaling law ($\alpha \sim 2/3$). This deviation arises for several reasons. First, our system differs from 3D, incompressible, isotropic hydrodynamics, as it incorporates plasma effects, anisotropies induced by shears, and reduced dimensionality. Second, it is important to emphasize that second-order statistics are not universal; they can only be approximated through scaling-law arguments and closures on time scales. The only exact law of turbulence pertains to third-order increments.

	Among the distinctions from classical isotropic hydrodynamics, it is important to emphasize that our weakly compressible systems exhibit anisotropic energy injection due to the magnetic field curvature $\zeta$. In this context, we present the two-dimensional structure function $S_2({\bf r})$ in the plane of 2D increments $(r_x, r_y)$, as shown in the right panel of Figure \ref{fig:S2}. 
	\comm{
	While the system displays many features of homogeneous isotropic turbulence, minor large-scale anisotropies are present due to poloidal shear flows along the $y$-coordinate – consistent with previous studies \cite{HidalgoEA03, SanchezEA09, Hidalgo2011, Halpern2016}. These lead to slightly different correlation lengths in the radial and poloidal directions and modest deviations in the $S_2$ scaling from classical isotropic turbulence. Aside from minor large-scale anisotropies induced by poloidal shear flows, the system demonstrates several characteristics of steady-state homogeneous isotropic turbulence, especially for $r\ll\lambda_c$.
	} Consequently, we will proceed to evaluate the third-order law in the next section.

	\begin{figure}
		\centering
		\includegraphics[width=0.9\linewidth]{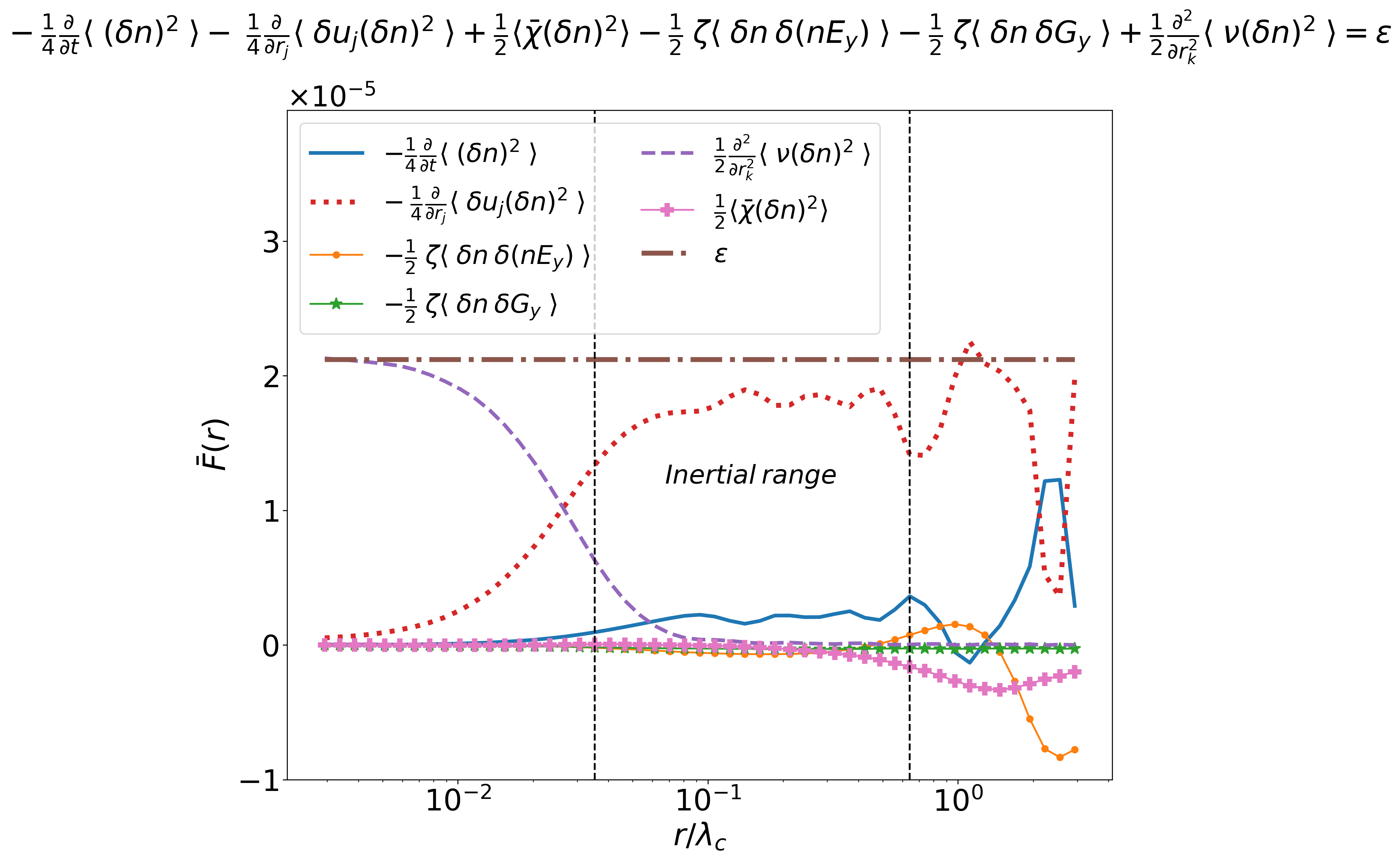}
		\caption{Yaglom-Braginskii law [Eq.~(\ref{final-law})] for direct numerical simulations of drift-electrostatic turbulence. Each term of the law, averaged over the polar coordinate as described by Eq.~(\ref{eq:polar}) \cite{WangEA2022}, is plotted as a function of the scale normalized to the correlation length.}
		\label{fig:Bra-Yaglom_RUN_II_bis}
	\end{figure}

	\section{\comm{The Yaglom-Braginskii Law in Simplified Simulations of Edge Turbulence}}
	The simulations of the drift-Braginskii equations reveal fully developed turbulence with clear homogeneity. The latter is the only fundamental requirement for the convergence of the statistics in Eq.~(\ref{final-law}). The third-order Braginskii-Yaglom law describes, scale by scale, the evolution, and distribution of density fluctuations. In analogy with the Kolmogorov 4/5 law \cite{KOLMOGOROV_41a, KOLMOGOROV_41b}, it can indicate which quantity dominates at a given scale, offering insights into the processes that primarily influence the fluctuations at that scale.

	Following a methodology analogous to that used for the second-order statistics in the previous section, we computed all seven terms in Eq.~(\ref{final-law}). These terms were evaluated on a polar grid $(r, \theta)$, where $\theta$ is the angle formed by the increment with the $\bf x$ direction in the $(r_x, r_y)$ plane. The use of a polar grid not only accelerates the computation but also facilitates the visualization of convergence. For instance, the main energy flux term was computed as the vector ${\bf Y}= -\frac{1}{4}\langle \delta {\bf u} (\delta n)^2\rangle$, where ${\bf Y} = (Y_r, Y_\theta)$. The averages were accumulated in physical (Cartesian) space using high-order bilinear interpolation, and the divergence $F(r, \theta)={\bf \nabla}_r \cdot {\bf Y}$. A grid with $N_r = 50$ radial points and $N_\theta = 32$ angular points was employed. All other terms, including scalar source terms, were computed similarly. After evaluating all terms, we performed a polar average following the approach suggested by Nie and Tanveer \cite{Nie-Tanveer1999, WangEA2022}. Specifically, for each term, we computed the polar average as
	\begin{gather}
		\overline{F}(r) = \frac{1}{2 \pi} \int_0^{2 \pi} F(r, \theta) \, d\theta, 
		\label{eq:polar}
	\end{gather}
	which provides highly accurate (and statistically relevant) measurements. The cascade law is shown in Figure \ref{fig:Bra-Yaglom_RUN_II_bis}, where it is evident that different terms dominate in distinct regimes.

	At very large scales, $r > \lambda_c$, the dominant term is typically the time variation of the second-order moment, $-(1/4)\frac{\partial }{\partial t}\left \langle (\delta n)^2 \right\rangle$. This term exhibits significant fluctuations, reflecting the temporal evolution of energy-containing scales. On average, it is positive, indicating energy injection from large scales. The shearing term due to magnetic curvature, $-(1/2)\zeta \left\langle \delta n \delta(n E_y)\right\rangle$, also contributes significantly at correlation scales, introducing anisotropy into the system. Similarly, the compressibility-driven source term, $(1/2)\left\langle \delta \bar{\chi} (\delta n)^2\right\rangle$, operates at these injection scales. In contrast, the compressive shearing term, $-(1/2)\zeta \left\langle \delta n \delta G_y\right\rangle$, has a negligible contribution across all scales.

	For scales with $r /\lambda_c < 1$, the time variation term becomes insignificant, and the dynamics are entirely dominated by the Yaglom flux term, ${\bf \nabla}_r \cdot {\bf Y}$. This behavior closely resembles that observed in hydrodynamic and MHD turbulence \cite{VerdiniEA15, SorrisoEA02, WanEA12, YangEA21, AdhikariEA23}, suggesting the existence of a well-defined range where the cross-scale transfer is the dominant process. This unambiguously identifies the inertial range of turbulence.

	\begin{figure} 
		\centering
		\includegraphics[width=0.7\linewidth]{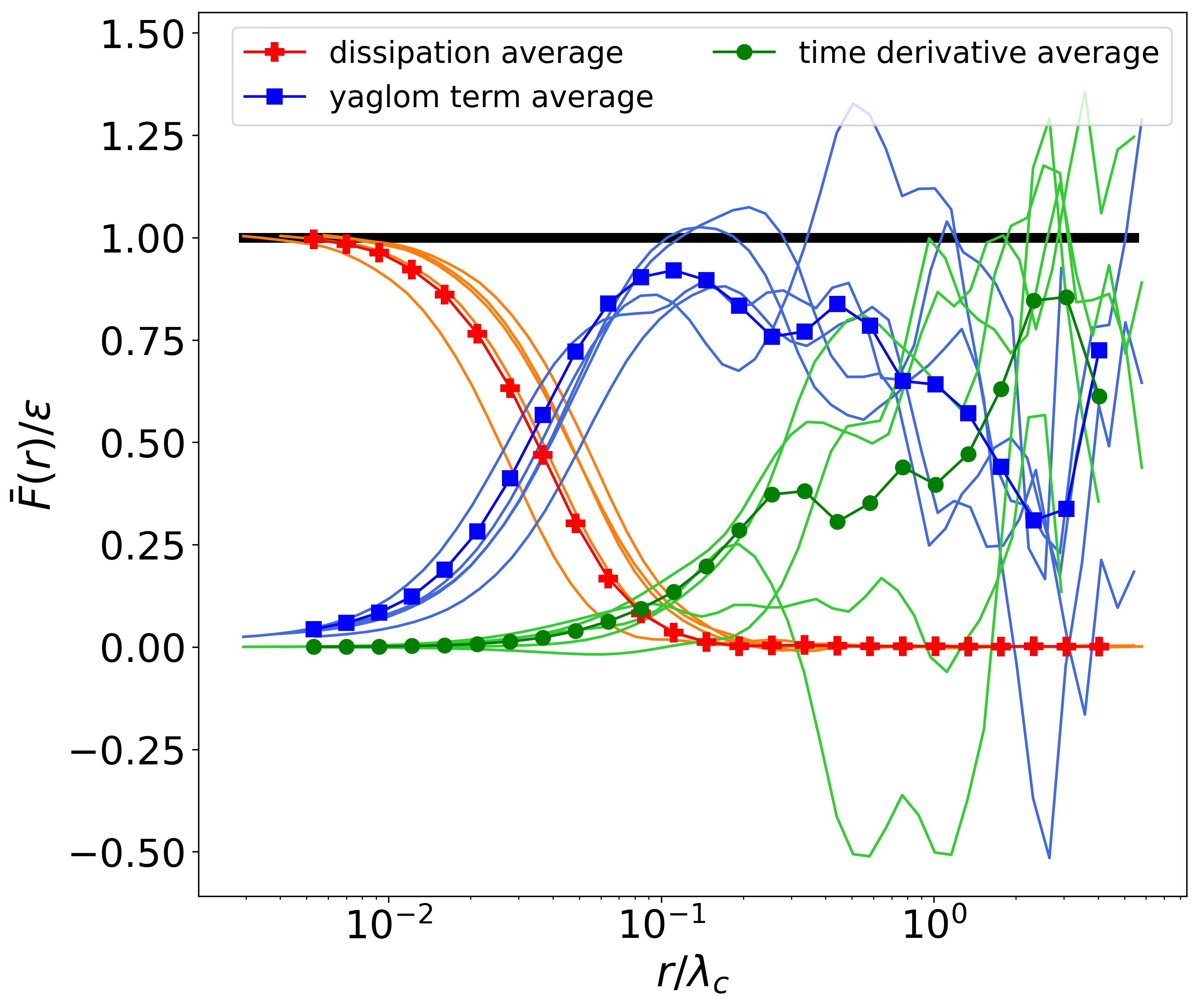}
		\caption{Comparison at different times of the main terms of the third-order law (lines). The time average is reported with lines and symbols. }
		\label{fig:Bra-Yaglom_average}
	\end{figure}

At very small scales, as $r \rightarrow 0$, the Yaglom term diminishes, and the dynamics are governed by the viscous contribution $\frac{1}{2}\frac{\partial^2}{\partial r_k^2} \left \langle \nu (\delta n)^2 \right \rangle$, where viscous dissipation suppresses the cascade. 
\comm{
	Note that a more physical treatment of such viscous contributions might change the extension of such dissipative regime, and will be investigated elsewhere.
}

On the same plot, we also include the total turbulence dissipation (or injection) rate $\varepsilon$. As shown, this rate matches different contributions depending on the regime. At small scales, it primarily reflects dissipation; in the inertial range, it aligns with the Yaglom flux; and at very large scales, it corresponds to a combination of time variability and various source terms.

The law holds throughout the simulation, and to enhance the statistical significance, we applied the above procedure at different times, and also by varying the initial random perturbations to the large scale equilibrium. After performing the analysis illustrated in Figure \ref{fig:Bra-Yaglom_RUN_II_bis}, we normalized all terms to the value of $\varepsilon$ at that specific time. Once these normalized terms were measured, as shown in Figure \ref{fig:Bra-Yaglom_average} (lines), we binned the contributions at given scales. A logarithmic spacing was used for the normalized increments $r/\lambda_c$. Finally, the time-averaged results are presented in the same figure (lines and symbols). The time-averaged law confirms the picture obtained from the single-time analysis: time variability dominates at large scales, while the Yaglom flux prevails at $r < \lambda_c$, diminishing at very small scales where viscous effects terminate the cascade.

	As shown in Figure \ref{fig:Bra-Yaglom_average}, an inertial range scaling is evident, particularly for $8 \times 10^{-2} < r/\lambda_c < 1$. There are alternative methods to highlight this range that are less stringent than computing the divergence, which is highly sensitive to convergence. Among these, a commonly used approach is to integrate the law over a given volume in the increments space within the inertial range \cite{MarinoEA2023, WanEA10, FerraroEA16 }. This procedure improves convergence but can be influenced by large-scale anisotropies near $\lambda_c$. For completeness, and in analogy with the Kolmogorov 4/5 and 4/3 laws \cite{Eyink02}, we assume the existence of a perfect cascade in the inertial range, where 
	\begin{gather}
		\varepsilon \simeq -\frac{1}{4} {\bf \nabla}_r \cdot \left \langle \delta {\bf u} (\delta n)^2\right \rangle \equiv  -\frac{1}{4} {\bf \nabla}_r \cdot  {\bf Y}.
	\end{gather}
	Assuming the inertial range is approximately isotropic, we integrate over a circular section, yielding
	\begin{gather}
		-4 \varepsilon \int_0^{2 \pi} \int_0^r r' \, dr' \, d\theta = \int_0^{2\pi}\int_0^r {\bf \nabla}_r \cdot  {\bf Y} \, r' \, dr' \, d\theta,  
	\end{gather}
	which directly leads to the 2D, direction-averaged, third-order law:
	\begin{gather}
		\overline{Y}_r(r) = - 2 \varepsilon r.
		\label{eq:polav}
	\end{gather}

	\begin{figure} 
		\centering
		\includegraphics[width=0.7\linewidth]{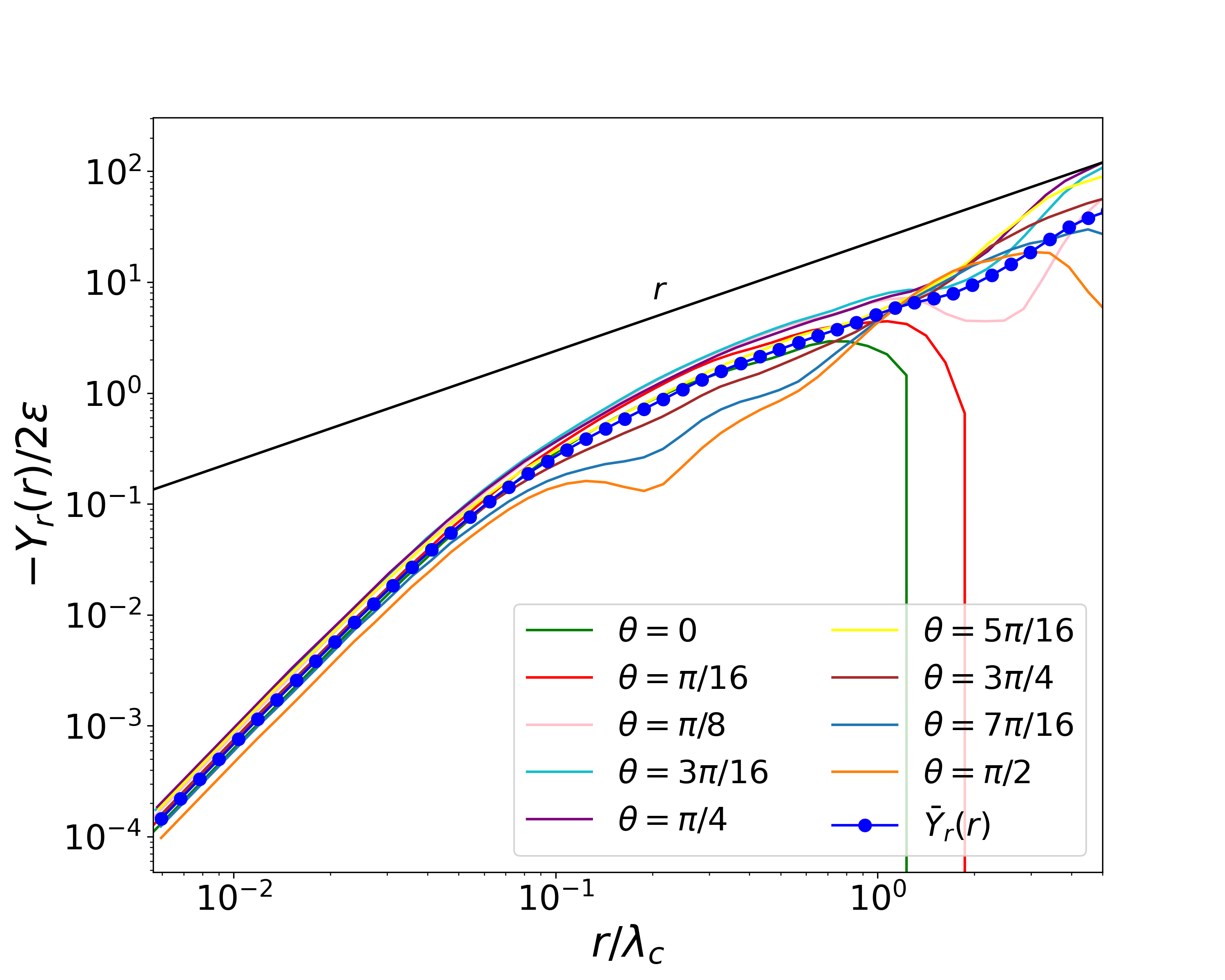}
		\caption{Radial component of the Yaglom flux $Y_r(r)$, nomalized via  $-2 \varepsilon$, at several angles (lines). The polar average (along $\theta$), computed via Eq.~(\ref{eq:polav}) is reported with thick (blue) line and symbols. }
		\label{fig:Yr_polar}
	\end{figure}

	In Figure \ref{fig:Yr_polar} we report the radial component of the Yaglom flux, at several polar angles, together with its average. The scaling law $\overline{Y}_r\sim r$ is even more clear in this form and has a broader extension but with higher uncertainty due to the anisotropy of the large scales.

	\section{Conclusions}
	\label{sec:conclusions}

\comm{
	Understanding turbulent transport is crucial for optimizing magnetic confinement in tokamak devices.
}
	In this study, we introduce a novel high-order turbulence law that provides a potential framework for describing the turbulent cascade in fusion plasmas. Drawing on the Yaglom-Monin formalism, we derive an exact mathematical relationship that characterizes the behavior of density fluctuations in strongly magnetized systems. Specifically, we obtain a third-order von K\'arm\'an-Howarth equation in its increment form, adapted here to the drift-electrostatic Braginskii theory.

\comm{
	Fluid models, primarily based on reduced Braginskii equations, offer a computationally efficient yet physically meaningful description of the outer region. Here, we take a first step toward constructing a theorem for high-order turbulence analysis (particularly third-order increments) within a fluid approach - though a similar methodology could be adapted for gyrofluid, gyrokinetic, and full Vlasov codes \cite{HellingerEA18}.
}

	Following the approach proposed by Politano and Pouquet \cite{Politano_Pouquet1995}, we focus on the spatial lags of macroscopic fluctuation quantities, particularly the time variation of the second-order structure function of the density (a similar law can be derived for the temperature). The new Yaglom-Braginskii law represents a balance of various terms, such as the time variation of the second-order moment, the Yaglom inertial range flux, and the small-scale viscous term. By defining a turbulence injection rate for the density, \comm{ $\varepsilon = \langle \nu |\nabla n|^2 \rangle$ }, in analogy with the passive tracer form of the Yaglom-Obukhov law, we also account for new source terms arising from large-scale magnetic shears and the compressibility of the system.

	To validate Eq.~\ref{final-law}, we performed direct numerical simulations using a simplified (reduced) model. These simulations were performed in a decaying turbulence regime, initiated by perturbing large-scale density and temperature profiles. After an initial transient, turbulence develops, and a quasi-steady-state regime is established, where energy is injected by large-scale shears, transferred to smaller scales, and eventually left the system through open boundaries. By selecting a region of homogeneous turbulence during this steady state, we measured all terms of the new law. 
	\comm{
		We observed that the law in Eq.(\ref{final-law}) consists mainly of a few main contributions that act at separate scales. At large scales, $r>\lambda_c$, the dynamics are dominated by injection mechanisms, including instabilities driven by large-scale gradients and contributions from magnetic field inhomogeneities and the electrostatic field. The intermediate scales exhibit a Yaglom-like energy transfer term, analogous to the universal cascade behavior observed in hydrodynamic passive scalar turbulence. Finally, at small scales, the cascade is terminated by the viscous (dissipative) terms, which might depend on the collisionality regime of the plasma. The picture is relevant in the context of edge and SOL turbulence.
	}

	\comm{
		The analysis demonstrates that the plasma dynamics maintain a consistent cross-scale balance across different configurations, exhibiting a well-defined turbulent inertial range. Additional simulations (not shown) confirmed that these findings remain valid when varying both the initial perturbation spectrum and amplitude, with the third-order scaling law being particularly robust to these changes.
    }
	This study highlights the effectiveness of the Yaglom-Braginskii third-order law in accurately measuring the cascade rate of density fluctuations.

	\comm{
		Future investigations will focus on both the implementation in full 3D geometry to examine anisotropic energy transfer, particularly along the parallel direction, as well as the inclusion in our reduced model of inner-boundary forcing and SOL loss terms to better replicate experimental conditions. While our current cross-scale balance (Eq.~\ref{final-law}) remains valid for non-stationary conditions, the transition to a steady-state would very likely eliminate the temporal derivative terms and introduce new boundary flux components.
	} 
These high-order analyses can be employed to extract the injection or dissipation rate of turbulence, either through multi-probe measurements or reduced one-dimensional measurements, under the assumption of local isotropic fluctuations \cite{SorrisoEA18, ServidioEA2022}. Such measurements are especially relevant in the SOL of tokamaks, a critical region for understanding plasma behavior in fusion experiments.

	\section*{Acknowledgements}
	The simulations have been performed at the Alarico cluster at the University of Calabria. S. S. acknowledge supercomputing resources and support from ICSC-Centro Nazionale di Ricerca in High Performance Computing, Big Data, and Quantum Computing–and hosting entity, funded by European Union-NextGenerationEU. Finally, the authors would like to acknowledge Prof. Vincenzo Carbone, who passed away in January 2025, for inspiring this work.

	\section*{References}
	\bibliography{iopart-num}

	\clearpage

\end{document}